\documentclass[10 pt,final,journal,letterpaper,oneside,onecolumn]{IEEEtran}
\usepackage{multicol}   
\usepackage{multirow}
\usepackage[pdftex]{graphicx}
\usepackage[lined,boxruled]{algorithm2e}
\usepackage{algorithmic}
\usepackage[small]{caption}
\usepackage{float}
\usepackage{xspace}
\usepackage{fancyhdr}
\usepackage{amssymb,amsmath}
\usepackage{subfig}
\usepackage{textcomp} 
\usepackage{epstopdf}
\usepackage{bbding}
\usepackage{color}
\usepackage[bottom]{footmisc}
\begin{document}

\title{Applications of Backscatter Communications for Healthcare Networks}

\author{Furqan Jameel, Ruifeng Duan, Zheng Chang, Aleksi Liljemark, Tapani Ristaniemi, Riku Jantti}

%\markboth{IEEE Wireless Communications}%
%{\MakeLowercase{\textit{et al.}}: Wireless Communications}

\maketitle

\begin{abstract}
Backscatter communication is expected to help in revitalizing the domain of healthcare through its myriad applications. From on-body sensors to in-body implants and miniature embeddable devices, there are many potential use cases that can leverage the miniature and low-powered nature of backscatter devices. However, the existing literature lacks a comprehensive study that provides a distilled review of the latest studies on backscatter communications from the healthcare perspective. {Thus, with the objective to promote the utility of backscatter communication in healthcare, this paper aims to identify specific applications of backscatter systems.} A detailed taxonomy of recent studies and gap analysis for future research directions are provided in this work. Finally, we conduct measurements at 590 MHz in different propagation environments with the in-house designed backscatter device. The link budget results show the promise of backscatter devices to communicate over large distances for indoor environments which demonstrates its potential in the healthcare system.
\end{abstract}


\IEEEpeerreviewmaketitle

\section{Introduction}
The medical industry today is seeking new solutions for in-body and on-body devices that transfer the data over a wireless channel \cite{yuce2012easy,chen2009wireless}. {This includes pacemakers for generating electric pulses, micro-scale robots that operate in the bloodstream, and smart pills for identifying abnormalities in the gastrointestinal tract.} However, the modern deep tissue systems consume a significant amount of energy by generating their own radio signals. For instance, wireless capsules for endoscopy consume up to 10 times more power than the sensors \cite{yuce2012easy}. {Due to these reasons, the large battery of the capsule consumes 40-50\% of the total space of the capsule \cite{chen2009wireless}. Reduction in this form-factor (i.e., the size, shape, and other physical specifications of electronic components) of these capsules can not only improve the likelihood of completion of endoscopy but also make them easy to swallow and excrete.} Similar challenges are faced in the case of on-body sensors. The premise of on-body sensor networks is to build a network of devices capable of operating in a battery-free manner by means of smart networking, and power management at the granularity of individual bits and instructions. This is challenging to achieve through conventional networking approaches due to the need for active radio circuits, large form-factors, and their energy constraint nature. Thus, we expect that it is important to divorce the healthcare from conventional wireless solutions and move towards innovative systems for seamlessly connecting the in-body and on-body wireless devices. 
%%%%%%%%%%%%%%%%%%%%%
\begin{figure*}[!htp]
\centering
\begin{tabular}{c}
\includegraphics[trim={0 8cm 0 0cm},clip,scale=.45]{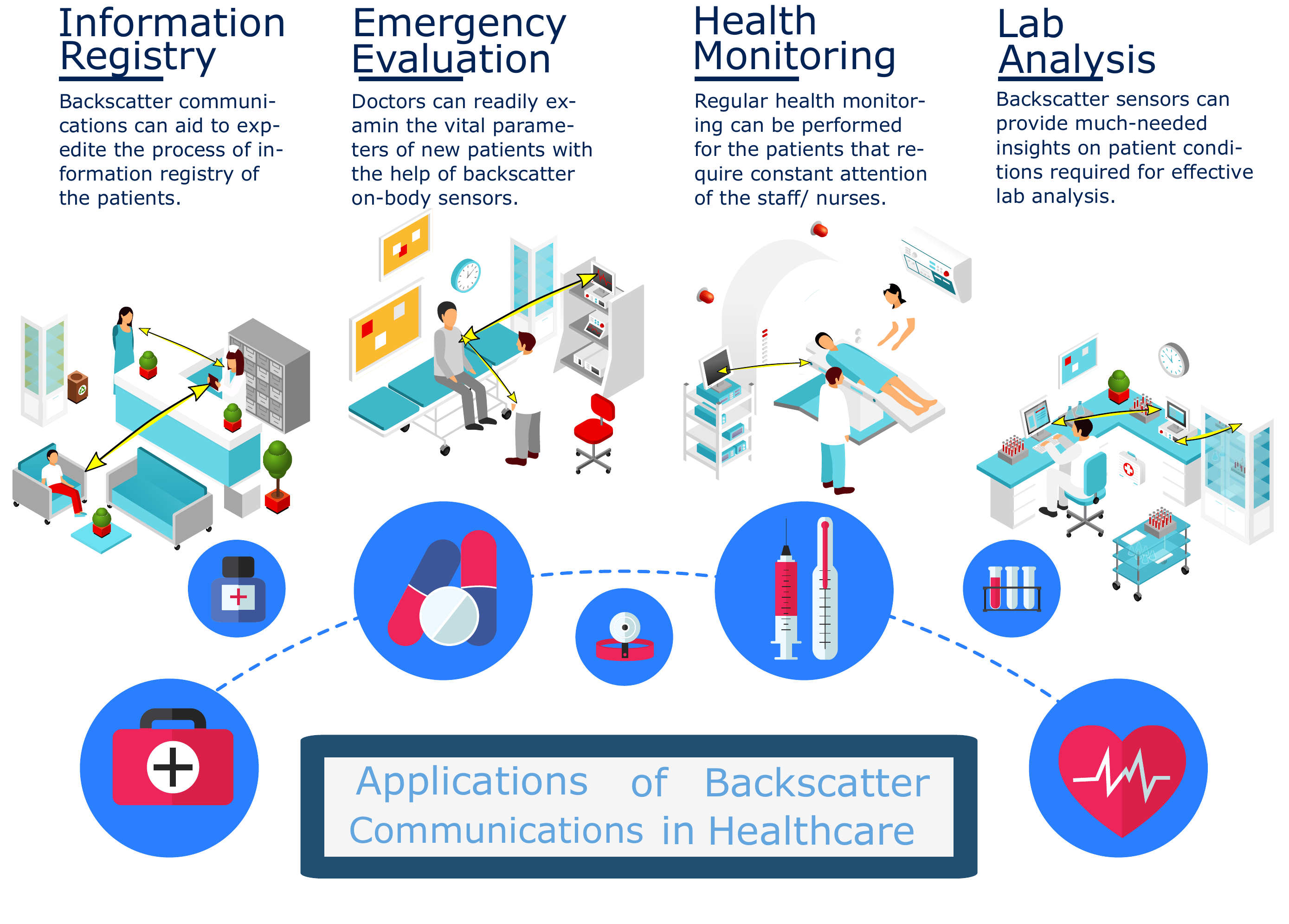}
\end{tabular}
\caption{{Applications of backscatter communications in healthcare. These applications include information registry that can aid to expedite the process of registration of the patients, evaluation of body conditions in the emergency room, regular health monitoring, assistance in lab analysis of patient's data.}}
\label{fig1}
\end{figure*}

Backscatter communication is an emerging paradigm and a key enabler for pervasive connectivity of low-powered wireless devices \cite{van2018ambient}. It is primarily beneficial in situations where computing and connectivity capabilities expand to sensors and miniature devices that exchange data on a low power budget. Due to this innovative method of communication, the backscatter communications provide virtually endless opportunities to connect wireless devices \cite{Duan}. Wearable devices, connected homes, industrial Internet of things (IoT), and miniature embeddable are some of the areas where backscatter communications can be used to provide pervasive connectivity. Besides this, backscatter devices can enable deep tissue localization for isolating abnormal tissues and depositing biomarkers at the location. In fact, we anticipate that the potential benefits of backscatter communications would significantly impact how doctors and patients interact with wireless devices. Some potential applications of backscatter communications in healthcare are depicted in Figure \ref{fig1}.

The user-focused perspective of backscatter communication makes it a perfect contender for the healthcare domain. As shown in Table \ref{my-table3}, its low-powered architecture is well-suited for not only capturing the sensory data but also transmitting it over several meters \cite{van2018ambient}. {Note that the values for backscatter communication in Table \ref{my-table3} can vary depending on the RF technology, hardware design and the coverage requirements of the system.} The transmission distance can be extended if appropriate backscatter communication techniques are used. Moreover, by efficiently exchanging the data under low power budget, it can completely alter the way healthcare service is provided through embedded and wearable devices \cite{ensworth2017ble}. For instance, it can enable a more personalized form of healthcare for self-monitoring and self-managing wireless sensor systems. The integration of backscatter communication solutions with office systems, community health services, and electronic health records can also be a prime mover for smart cities. {In short, backscatter communication has the potential to increase the longevity of wearables and on-body sensor networks without compromising the form-factors of the individual devices.}

{The aforementioned aspects motivate us to further explore the utility of backscatter communications in healthcare.} To accomplish this objective, we first explain the fundamentals of backscatter communication systems. Then, we review a few reported recent studies on applications of backscatter communication in healthcare. Next, based on these studies, we provide a gap analysis through an extensive taxonomy. Then, we conduct a comprehensive study for in-house designed backscatter device that operates on 590 MHz. We perform measurements for indoor propagation conditions and provide link budget analysis. Finally, we provide future research directions for effectively employing backscatter communications in healthcare services.  
%%%%%%%%%%%%%%%%%%%%%%%
\begin{table*}[!htp]
\caption{{Comparison of different technologies with backscatter communications.}}
\label{my-table3}
\centering
\begin{tabular}{|p{2cm}|p{1.5cm}|p{1cm}|p{1.5cm}|p{1.5cm}|p{1.5cm}|p{1.5cm}|p{1.5cm}|}
\hline
\textbf{Standard/ Technology} & \textbf{Zigbee} & \textbf{ANT} & 
\textbf{Bluetooth 4.0} & \textbf{RuBee} & \textbf{Z-Wave} & \textbf{Insteon} & \textbf{Backscatter Comm.}\\
\hline \hline
Coverage area (meter)& 30 - 100 & Average House & 10 & 30 & 30 & Average House & 1 - 500 \\ \hline
Data rate (Kb/sec)& 250 & 1000 & 3000 & 9.6 & 9.6 & 13 & 10 - 1000 \\ \hline
Frequency band (MHz)& ISM & 2400 & 2400 & 0.131 & 900 & 902 - 924 & 512 - 2400  \\ \hline
Network topology& Mesh & Star & Star & Peer-to-peer (P2P) & Mesh & Mesh & P2P or Cooperative\\ \hline 
\end{tabular}
\end{table*}
%%%%%%%%%%%%%%%%%%%%%

\section{Fundamentals of Backscatter Communications} 

A standard backscatter system consists of a backscatter transmitter, a receiver, and a carrier emitter. The backscatter transmitter reflects the signal from the emitter to the receiver \cite{lu2018ambient}. On the other end, the receiver is designed in such a way that it is capable of decoding the modulated signal from the backscatter transmitter. On the basis of operation, the backscatter systems can be classified into passive and semi-passive communication systems. In the case of passive communications, the backscatter device relies on the harvested energy from the incident RF waves \cite{assimonis2016sensitive}. The harvested energy is then used to activate the device. By contrast, the semi-passive backscatter devices are equipped with an internal power source. {This availability of the power source ensures small access delays and improved reliability.}    

On the basis of carrier emitter, the backscatter devices can be configured in one of three ways, i.e., monostatic, bistatic, and ambient. In the monostatic configuration, the carrier emitter is installed inside the interrogator and a backscatter device reflects the signal upon a query \cite{lu2018ambient}. A bistatic configuration consists of a dedicated carrier emitter, backscatter transmitter, and a backscatter receiver \cite{Duan}. The backscatter transmitter uses the RF waves from the carrier emitter to forward the message to the backscatter receiver. The studies have shown the superiority of bistatic configuration over conventional monostatic configuration both in terms of cost-effectiveness and reliability \cite{kimionis2014increased}. Finally, ambient backscatter configuration works on the principle of exploiting ambient RF waves as carrier emitter. The backscatter transmitter uses the existing RF waves to reflect the message to the receiver. These ambient RF sources can be nearby base stations, TV/ FM towers, Bluetooth or Wi-Fi access points \cite{ensworth2017ble}.  

Among the aforementioned configurations of backscatter communications, the ambient backscatter communication is perhaps the most energy and spectrum efficient. First of all, there is no requirement of a dedicated carrier emitter in ambient backscatter configuration. The ambient backscatter devices operate using the carrier waves of the ambient RF sources like TV/ FM stations and Wi-Fi. This saves the energy consumed by a dedicated carrier emitter \cite{van2018ambient}. Secondly, by exploiting the existing RF signals, the ambient backscatter configuration becomes spectral efficient as it does not require extra spectrum to operate. {Moreover, the ambient backscatter configuration does not result in noticeable interference unless the devices are placed very close \cite{van2018ambient}.} Finally, the low cost and small form-factor of ambient backscatter devices favor its large-scale deployment in myriad scenarios. From healthcare to home automation, the ambient devices can be successfully used to complete a number of tasks at a very low cost.   

{Prior to shedding light on the wireless-powered ambient backscatter communications, it is worthwhile to report the basics of wireless power transmission. Wireless-powered communication has recently emerged as an active paradigm to increase the lifetime of devices.} The rapid developments in hardware sensitivity, minimization of circuit power consumption, and improved RF-to-DC conversion solutions have paved the way for ambient RF energy harvesting. {The combination of wireless power transmission and ambient backscatter communication has shown quite promising results and holds the key to self-sustainable healthcare services. Additionally, with the integration of wireless power transfer, the backscatter communications can inspire new ways of recharging existing wearable devices, e.g., Google Glass, Pebble smartwatch, and Fitbit health monitor \cite{metcalf2016wearables}.} 

\section{Recent Advances in Healthcare: A Backscatter Communication Perspective}

In this section, we highlight, investigate, and report recent advances in healthcare domain from the backscatter communications perspective. Based on the evaluation of these studies, we devise a taxonomy in the next section by classifying and categorizing the backscatter literature.

{Vasisht \emph{et al.} introduce a novel backscatter design for deep tissue devices in \cite{vasisht2018body}.} They identify that deep tissue devices suffer from two major issues, i.e., signal deflection and surface interference. The signal deflection is caused by the fact that the RF signal travels 8 times slower in muscles than in air. Due to this difference, most of the signal is reflected and the implants experience exponentially more attenuation in human muscles as compared to air. To address these challenges, they propose a backscatter communication design called ReMix that allows non-linear mixing of RF signals. They have shown that for devices that are 8 cm deep in the tissues of the animal, ReMix achieves up to 17 dB of signal-to-noise ratio (SNR) for the bandwidth of 1 MHz. They also show that the proposed design can locate deep-tissue implant up to an average accuracy of 1.4 cm.

In the domain of multi-hop communications, the authors of \cite{zhao2018x} present a multi-hop backscatter communications system called X-Tandem. They consider indoor monitoring and on-body sensing tags as some of the potential applications of X-Tandem. The design of X-Tandem utilize the waves generated from off-the-shelf Wi-Fi as an ambient RF signal. The communication range is extended by relaying the information across multiple backscatter devices, wherein, the backscattered packet is constructed as a legit Wi-Fi packet. The experimental results demonstrate that X-Tandem achieves the throughput of 200 bps which is sufficient for collecting sensor data. {However, as the distance increases up to 6.5 m the throughput drops down to only 50 bps.} In a similar way, the authors of \cite{wang2017fm} explored a novel FM-based backscatter architecture for smart fabric. The performance of the system was evaluated by transmitting data at 1.6 kbps and 100 bps in an outdoor environment. It was found that the prototype was able to receive -35 dBm to -40 dBm of radio signal level. {Moreover, the bit-error-rate (BER), while standing, was 0.2 for 1.6 kbps bit rate, whereas, it was only 0.005 at 100 bps even when the user was running.}     

Since both smart contact lens and patch sensors are worn on the body, their performance significantly degrades due to poor link quality and limited duty cycle. To address these issues, the authors of \cite{talla2017lora} used backscatter communications to enable reliable communication in smart contact lens and epidermal patch sensor. The authors leverage long-range (LoRa) backscatter communication to enable communication in contact lens and patch sensors. In case of smart contact lens, the backscatter device transmits LoRa packets at 1 MHz across a 104 ft by 32 ft atrium. It was shown that reliable connectivity can be ensured for received signal strength above -132 dBm. In a similar manner \cite{iyer2017inter}, the authors proposed an inter-technology backscatter solution for contact lens and implanted neural devices. The prototype is composed of 30 AWG wire antenna having 1 cm diameter. To maintain the structural integrity and biocompatibility, the antenna is encapsulated in a thin layer of polydimethylsiloxane (PDMS). The received power at Wi-Fi reaches up to -74 dBm when the distance between prototype and Wi-Fi is 10 inches, while for a Bluetooth receiver, the same range reaches up to -82 dBm. 

The authors of \cite{zhang2016enabling} design and implement a backscatter communications system for on-body sensors. Specifically, they designed a frequency-shifted (FS) backscatter system capable of communicating with commodity Wi-Fi and Bluetooth. The FS backscatter transmitter mainly consists of three major subsystems, i.e., clock generator, transmission logic, and RF transistor. Moreover, it was able to communicate over a longer distance (i.e., 4.8 m) as compared to conventional backscatter systems. The FS backscatter design was also suitable to bring a shift of 20 MHz to communicate with Bluetooth radios and commercial Wi-Fi. {Finally, the low power budget of FS backscatter (by using ring oscillator) allowed it to consume only 45 $\mu$W of power.} This low-powered nature of FS backscatter has made it more favorable for backscatter on-body sensors.

\section{Taxonomy of Backscatter Communications in Healthcare}

\begin{figure*}[!htp]
\centering
\begin{tabular}{c}
\includegraphics[trim={0 0cm 0 1cm},clip,scale=.6]{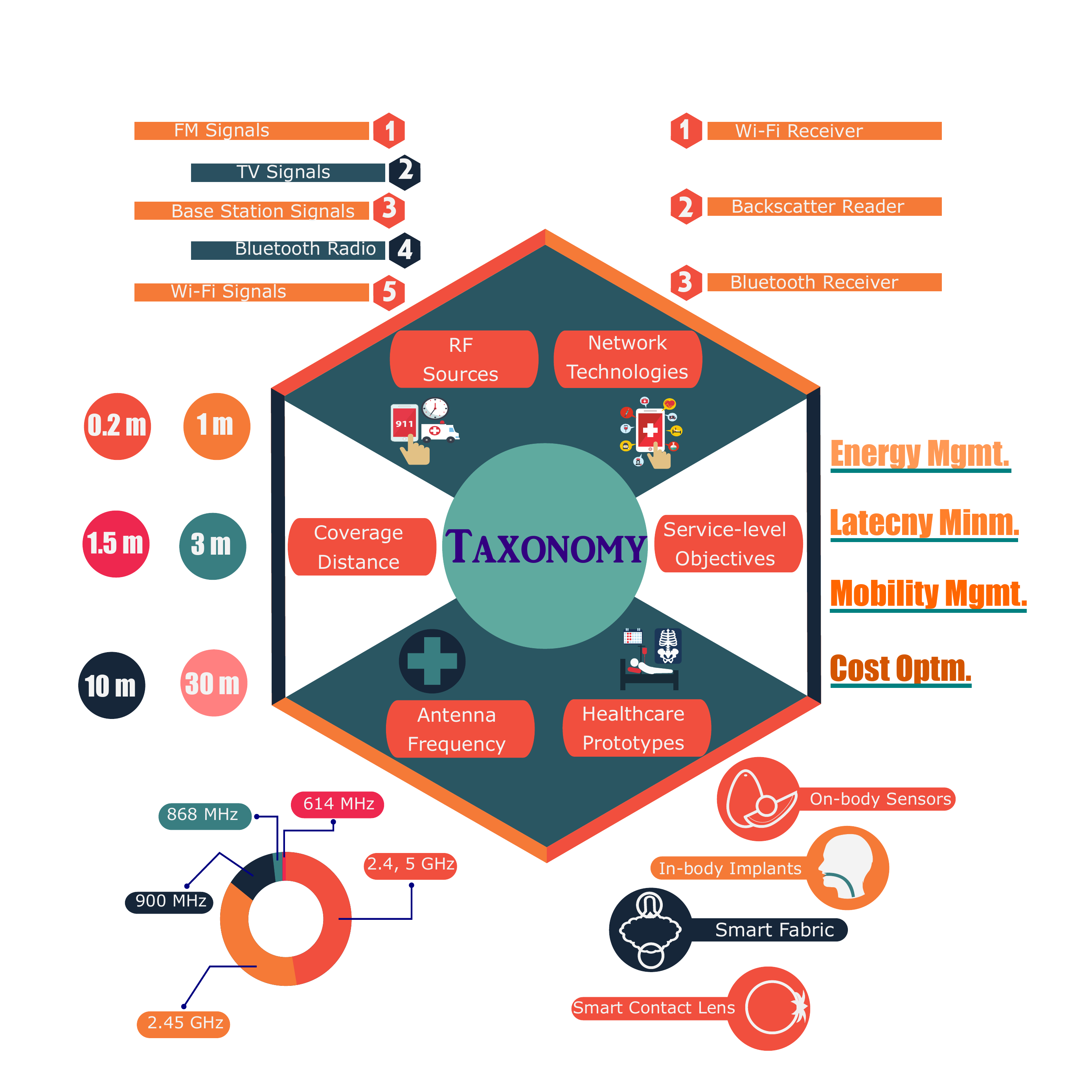}
\end{tabular}
\caption{Taxonomy of different features including network technologies, RF source, service-level objectives, antenna frequency, coverage distance, prototypes for backscatter communications in healthcare. {The network technologies for backscatter readers/ tags include Wi-Fi receivers and Bluetooth radios. Mobility considerations for fulfilling service-level objectives consider walking (i.e., 1.1 m/s) and running (i.e., 2.2 m/s) scenarios. While most of the backscatter applications are focused on conventional transmission of the sensed data to a gateway.}}
\label{fig2}
\end{figure*}

In light of the studies discussed above, a detailed taxonomy of healthcare applications of backscatter communication has been illustrated in Figure \ref{fig2}. The taxonomy is based on features like network technologies, RF source, service-level objectives, antenna type, coverage distance, and prototypes for healthcare applications.

%%%%%%%%%%%%%%%%%%%%%

\subsection{Network Technologies}

In the domain of healthcare, backscatter devices collect data from the sensor and transmit it to a central entity which is a receiver. The network technologies for backscatter communication can vary in terms of transmission range, achievable data rate, the total number of supported devices. The technologies range from backscatter readers/ tags to Wi-Fi receivers to Bluetooth radios. As per the analysis of recent studies, Wi-Fi receivers have proven to achieve the highest data rates due to their accessibility to multiple antennas and availability of free 12th and 13th channel. 

\subsection{RF Source}

The performance of backscatter devices largely depends on the type of RF source. {For indoor scenarios, RF signals from Wi-Fi and Bluetooth radios are a better choice, whereas, in the case of outdoor backscatter communication, RF signals from base stations, and FM and TV Towers are more preferred.} Due to high path loss and unavailability of line-of-sight in the indoor conditions, usage of FM and TV signals are very limited for healthcare applications. Moreover, the antenna form-factor is also not suitable for healthcare applications at lower frequencies (e.g., 900 MHz). It is, therefore, more appropriate to use RF source (like Wi-Fi and Bluetooth) with higher frequencies.   

\subsection{Service-level Objectives}

{Following are some of the service-level objectives that must be met in the context of healthcare applications:}

\subsubsection{Energy Management}

An important objective of backscatter devices is the efficient management of energy. Especially, in the presence of sensing platforms, the backscatter devices must manage energy at the granularity of individual bits and operations. Fortunately, in the case of healthcare applications, this energy management is much easier for indoor environments like hospitals. Fulfilling this key objective would make backscatter devices truly self-sustainable. 

\subsubsection{Latency Minimization}

High latency becomes a critical issue in the domain of healthcare. Since backscatter devices would monitor the vital parameters of a patient in real-time, timely delivery of packets is going to be required to ensure minimum latency. Multi-hop backscatter communication would also play a major role in maintaining the balance between reliability and latency.

\subsubsection{Mobility Management}

Backscatter solutions are more focused on the static conditions due to high power loss involved during mobility. {Although the reported studies consider walking (i.e., 1.1 m/s) and running (i.e., 2.2 m/s) scenarios, there is still room to explore mobility-aware backscatter communication techniques. So far, it has become evident that the data rates and the received signal strength drop significantly when the mobility of the user is considered.} This also impacts the throughput of the system which drops as little as 12 kbps in the worst-case as compared to 38 kbps for the same in a static condition. 

\subsubsection{Cost Optimization}

Backscatter solutions have been the most economical in terms of deployment and manufacturing cost. The miniature circuits are easy to construct and consume a very small amount of power. However, there is still room for improvement. Network node placement and planning are key determinants of cost. Optimal arrangement of RF sources and deployment of an optimal number of nodes can be considerably helpful in reducing the capital.  

\subsection{Antenna Frequency}

{Antenna type is mostly dependent on the RF source and impedance matching circuitry of the backscatter devices.}  For instance, consider Wi-Fi, Zigbee, and Bluetooth. Since these technologies share the same spectrum (from 2.4 GHz to 2.483 GHz), it is possible for a backscatter device to reflect a combination of these frequencies. However, it becomes very difficult for the same antenna (and impedance matching circuit) to reflect lower frequencies like 900 MHz. In this case, a more complex and sensitive receiver design would be required to decode the messages on all the frequencies. 
 
\subsection{Coverage Distance}

Coverage is mostly associated with large-scale deployment of backscatter devices. Efficient and reliable communication over large distances is most desirable in a healthcare application. Selection of the appropriate coverage distance includes estimation of energy resources at each device, coordination among devices, and proper allocation of the workload in the network.

\subsection{Healthcare Prototypes}

The delay sensitive nature of healthcare applications forces backscatter devices to react in a timely manner. For this reason, most of the backscatter applications are focused on transferring the sensed data to a receiver or gateway. Few have also considered the application in a smart contact lens. Still, the most prominent and emerging area of backscatter communication in health deals with in-body implants. The initial results have shown promise in terms of improved received signal strength and localization of in-body implants.

\section{Performance Evaluation}

Although recent studies in Section III have been instrumental in advancing the state-of-the-art, there still remain unresolved issues that need attention. {One of these issues pertains to the relationship between path loss, received signal strength and optimal distance of backscatter device. Since the received signal strength roughly determines the reliability of received messages, our study seeks to answer how far the RF source and the receiver be placed from the backscatter device in an indoor environment (like a ward or a hospital). With this intent, we develop an in-house miniature backscatter device, operating under 590 MHz, and evaluate its performance at different distances. From the healthcare perspective, our miniature backscatter device can be considered to be attached to a patient who moves freely inside a hospital. As the backscatter device moves in between RF source and the receiver, the received power provides us with an estimate of the reliability of communication when a patient is equipped with the device.} Moreover, we perform extensive simulations in MATLAB to confirm the measurement results based on the mathematical model\footnote{The details of mathematical derivations can be found in \cite{Duan}.}. 

%%%%%%%%%%%%%%%%%%%%%
\begin{figure*}[!htp]
\centering
\begin{tabular}{c}
\includegraphics[trim={0 0cm 0 0cm},clip,scale=.12]{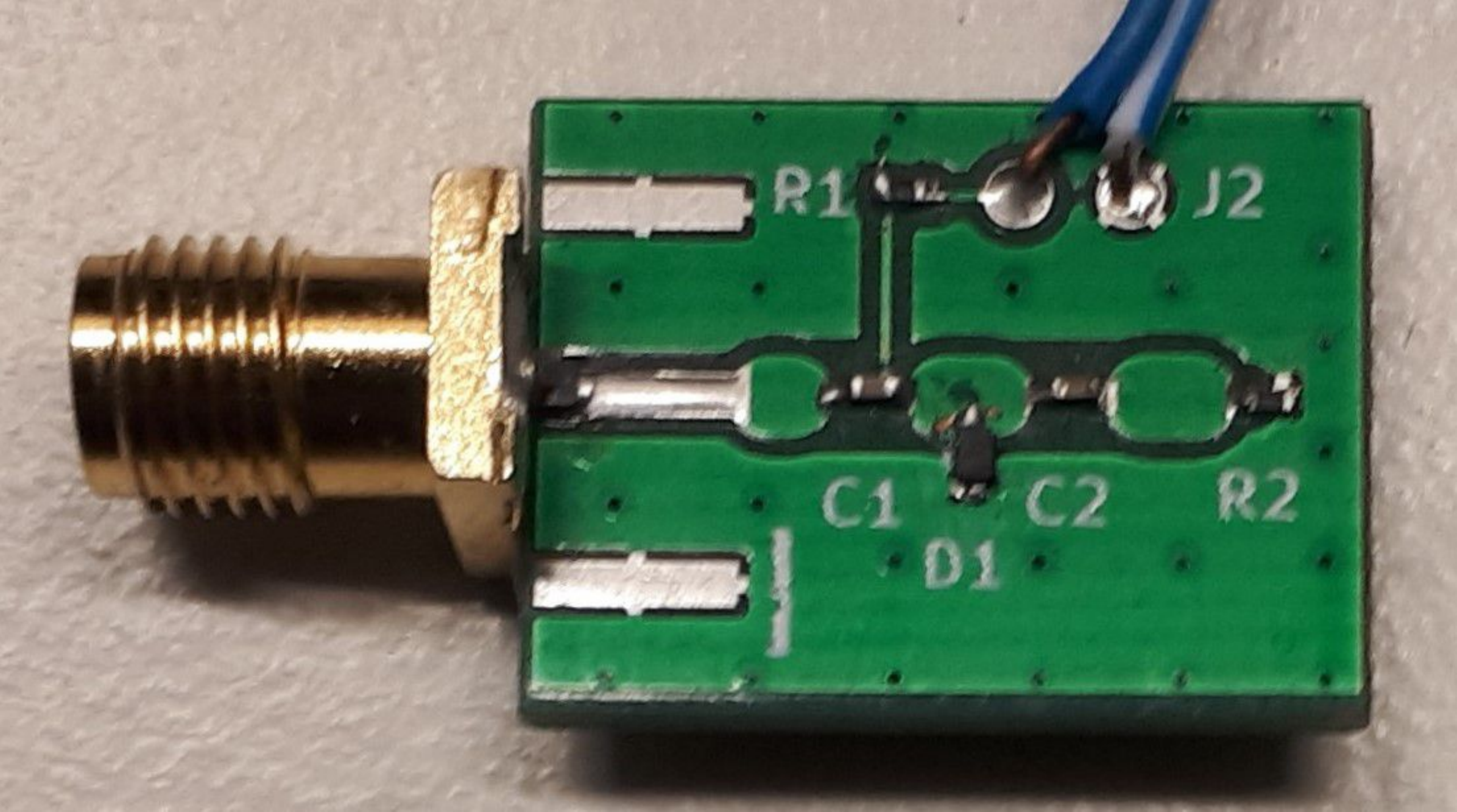} \\ 
(a) \\
\includegraphics[trim={0 0cm 0 0cm},clip,scale=.2]{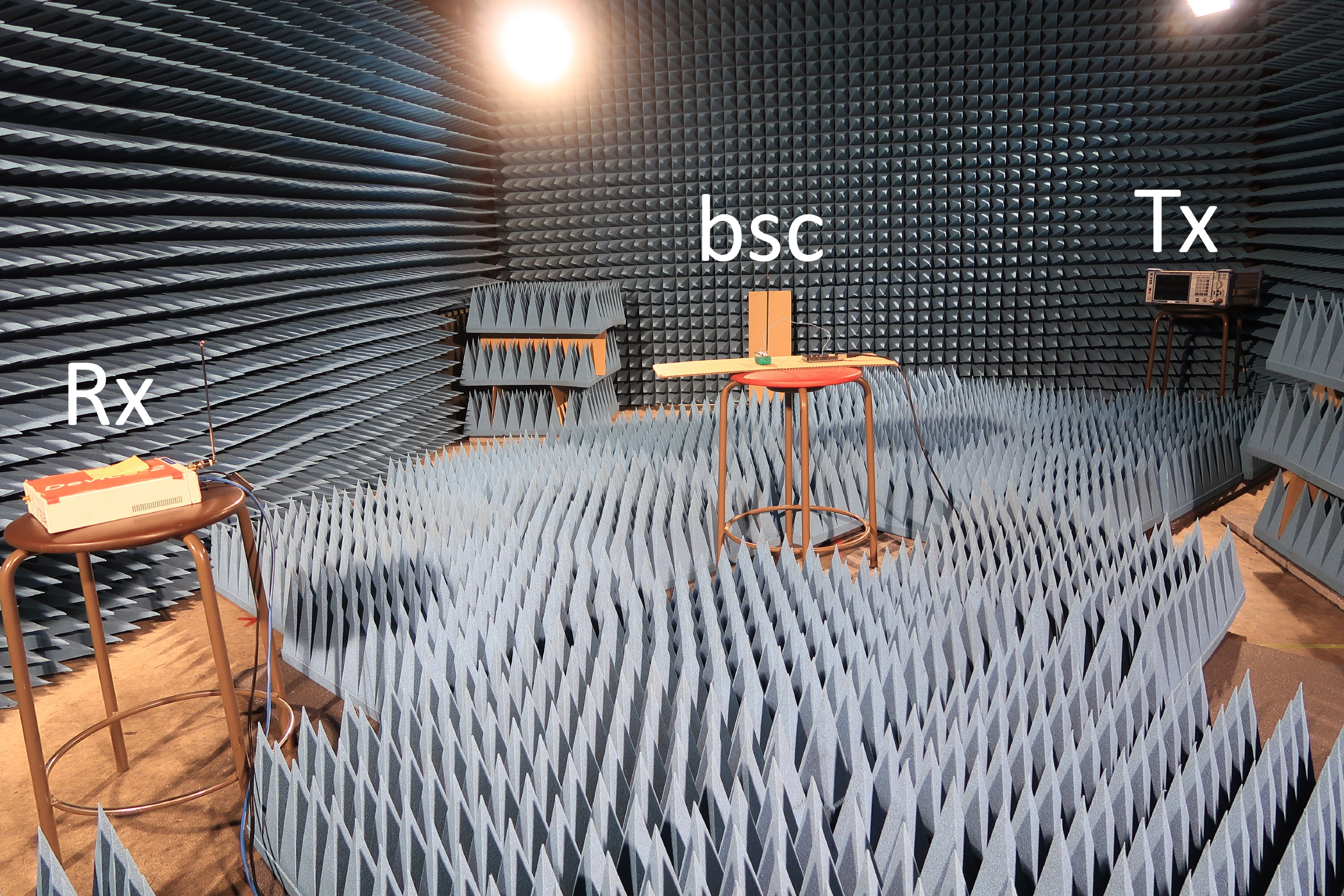} \\
(b) \\ 
\includegraphics[trim={0 0cm 0 0cm},clip,scale=.2]{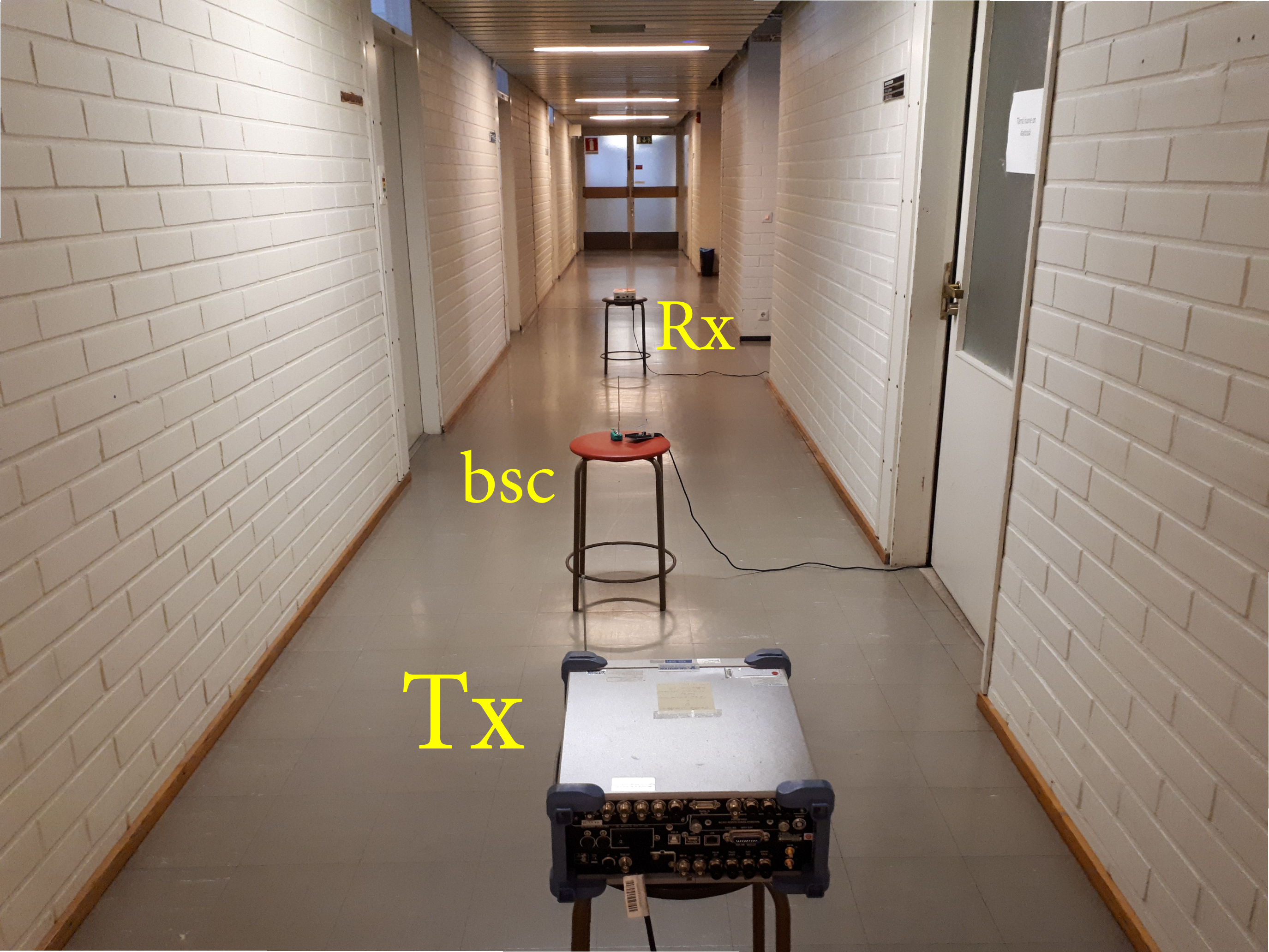} \\
(c)
\end{tabular}
\caption{{Experiment setup (a) In-house designed backscatter modulator and the configuration of resistors and capacitors (b) Anechoic chamber (c) Indoor corridor.}}
\label{fig3}
\end{figure*}

\subsection{Backscatter Design and Experiment Setup}

In this section, we provide the backscatter design and results for measurements performed in an anechoic chamber and indoor corridor. As illustrated in Fig. \ref{fig3}(a), we develop a backscatter modulator that uses on-off keying modulation. This modulation technique works by switching the carrier frequency on and off, as per the transferred binary message. This means that the receiver has to estimate if the ambient signal is either being reflected or not. The backscatter modulator is a small-scale modulator with 50 $\Omega$ resistor which represents a non-reflecting state. The capacitors C1 and C2 block the voltage to enter in terminating resistor or antenna. 

\begin{table}[]
\centering
\caption{Measurement parameters and their values.}
\label{tab2}
\begin{tabular}{|p{4cm}|p{3cm}|p{3cm}|}
\hline
\textbf{Parameters}          & \textbf{\begin{tabular}[c]{@{}l@{}}Value\\ (Corridor)\end{tabular}} & \textbf{\begin{tabular}[c]{@{}l@{}}Value\\ (Anechoic Chamber)\end{tabular}} \\ \hline
\multicolumn{3}{|l|}{\textbf{At RF Transmitter}}                                                                                                                                 \\ \hline
Transmit power               & 10 dBm                                                              & 10 dBm                                                                      \\ \hline
Number of antennas           & 1                                                                   & 1                                                                           \\ \hline
Type of signal               & Complex exponential                                                 & Constant carrier                                                            \\ \hline
Number of signal/ tones      & 1                                                                   & 1                                                                           \\ \hline
Frequency of operation       & 590 MHz                                                             & 590 MHz                                                                     \\ \hline
\multicolumn{3}{|l|}{\textbf{At Backscatter Device}}                                                                                                                             \\ \hline
Reflection coefficient (OFF) & -26.871 dBm                                                         & -26.871                                                                     \\ \hline
Reflection coefficient (ON)  & -0.2961 dBm                                                         & -0.2961                                                                     \\ \hline
Number of antennas           & 1                                                                   & 1                                                                           \\ \hline
Type of signal               & on-off keying wave (chip length 14 $\mu$s)                          & on-off keying wave (chip length 14 $\mu$s)                                  \\ \hline
\multicolumn{3}{|l|}{\textbf{At Receiver}}                                                                                                                                       \\ \hline
Number of antennas           & 1                                                                   & 1                                                                           \\ \hline
Sampling frequency           & 1 MHz                                                               & 1 MHz                                                                       \\ \hline
\multicolumn{3}{|l|}{\textbf{Misc.}}                                                                                                                                             \\ \hline
Power of received noise      & -107.4 dBm                                                          & -105.8                                                                      \\ \hline
Polarization                 & Vertical (whip antennas)                                            & Vertical (whip antennas)                                                    \\ \hline
Room temperature             & 296.15 K                                                            & 302.15 K                                                                    \\ \hline
\end{tabular}
\end{table}

For the experiment, an omnidirectional antenna with 0 dBi antenna gain is connected to the SMBV100A Rohde \& Schwarz signal generator. The operation frequency is 590 MHz and transmits power is 10 dBm. For the receiver to estimate whether the signal has been reflected or not, there must be a very large separation between the two transmission coefficients. We set the transmission coefficient for modulation as -0.3 dB while the non-modulation coefficient is set at -26.87 dB. Moreover, a software defined radio USRP-2932 is connected with a 0 dBi antenna gain for the reception of the signal. The software-defined radio is connected to a laptop which performs the post-processing of the signal. The height of the backscatter device and the receiver is 1 m, whereas, the height of the RF transmitter is 1.06 m.

In addition, we use a current controlled switch which is realized by using a diode. The diode requires a constant current of 10 mA. When the switch is not on, the diode is biased backward. This can be done by applying a negative voltage to the diode which results in minimizing the off-state capacitance. To do so, we derive 10 mA control current from a +5 V output pin of the microprocessor and apply it to the diode. It is worth pointing out that our design of the backscatter modulator does not limit the operating frequency. Rather, the limiting factor is the hardware quality of the components used to make the modulator. The termination resistor is perfectly matched to the transmission line in case of ideal components. In this state, the reflection coefficient is zero. However, the switching diode used for the experiment has 0.6 $\Omega$ forward resistance in the on state and a parasitic capacitance of 0.28 pF. At 590 MHz, the parasitic capacitance of 0.28 pF corresponds to capacitive reactance of 963 $\Omega$. Since this reactance is in parallel with the 50 $\Omega$ termination resistor, therefore, during non-reflecting state it gives a return loss of 31.7 dB. Moreover, the return loss in reflecting state is 0.21 dB which is determined by the forward resistance of the diode. Correspondingly, other hardware imperfections have a smaller impact on the performance of the backscatter modulator.

As shown in Fig. \ref{fig3}(b) and (c), the measurements have been performed for two indoor environments, i.e., anechoic chamber and office corridor. For both scenarios, the measurements are performed for dislocated configuration (i.e., when backscatter device and the receiver are separate) and we use the same equipment for all measurements. Due to the limitation of space in the anechoic chamber, the distance between the backscatter device and the receiver was varied up to 4 m while the backscatter device was placed between RF transmitter and receiver in a straight line. Whereas, in the case of the office corridor, the distance between the backscatter device and the receiver was varied up to 8 m. Table \ref{tab2} shows the measurement parameters for both indoor environments.

%%%%%%%%%%%%%%%%%%%%%
\begin{figure*}[!htp]
\centering
\begin{tabular}{cc}
\includegraphics[trim={0 0cm 0 0cm},clip,scale=.55]{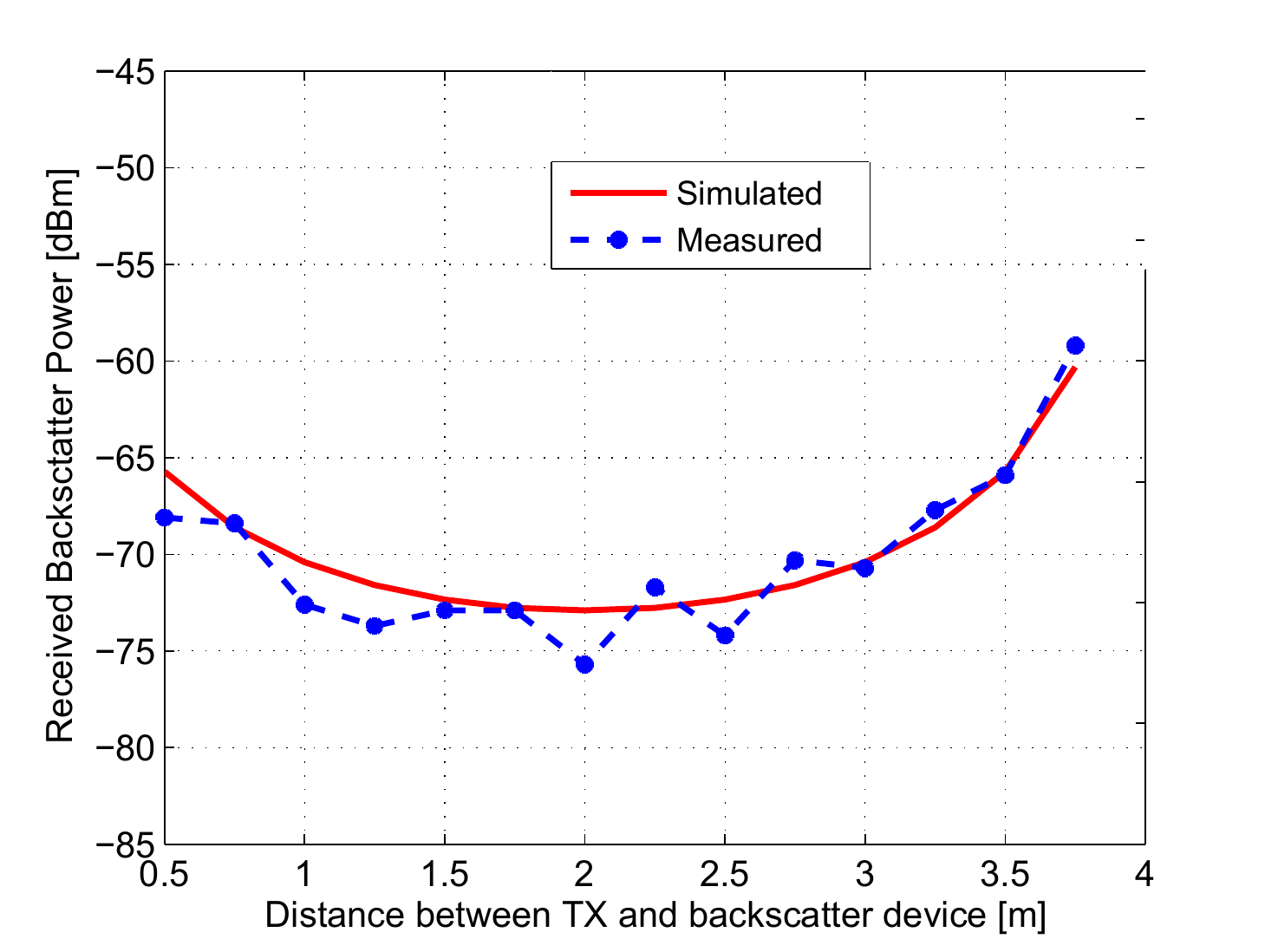} & \includegraphics[trim={0 0cm 0 0cm},clip,scale=.55]{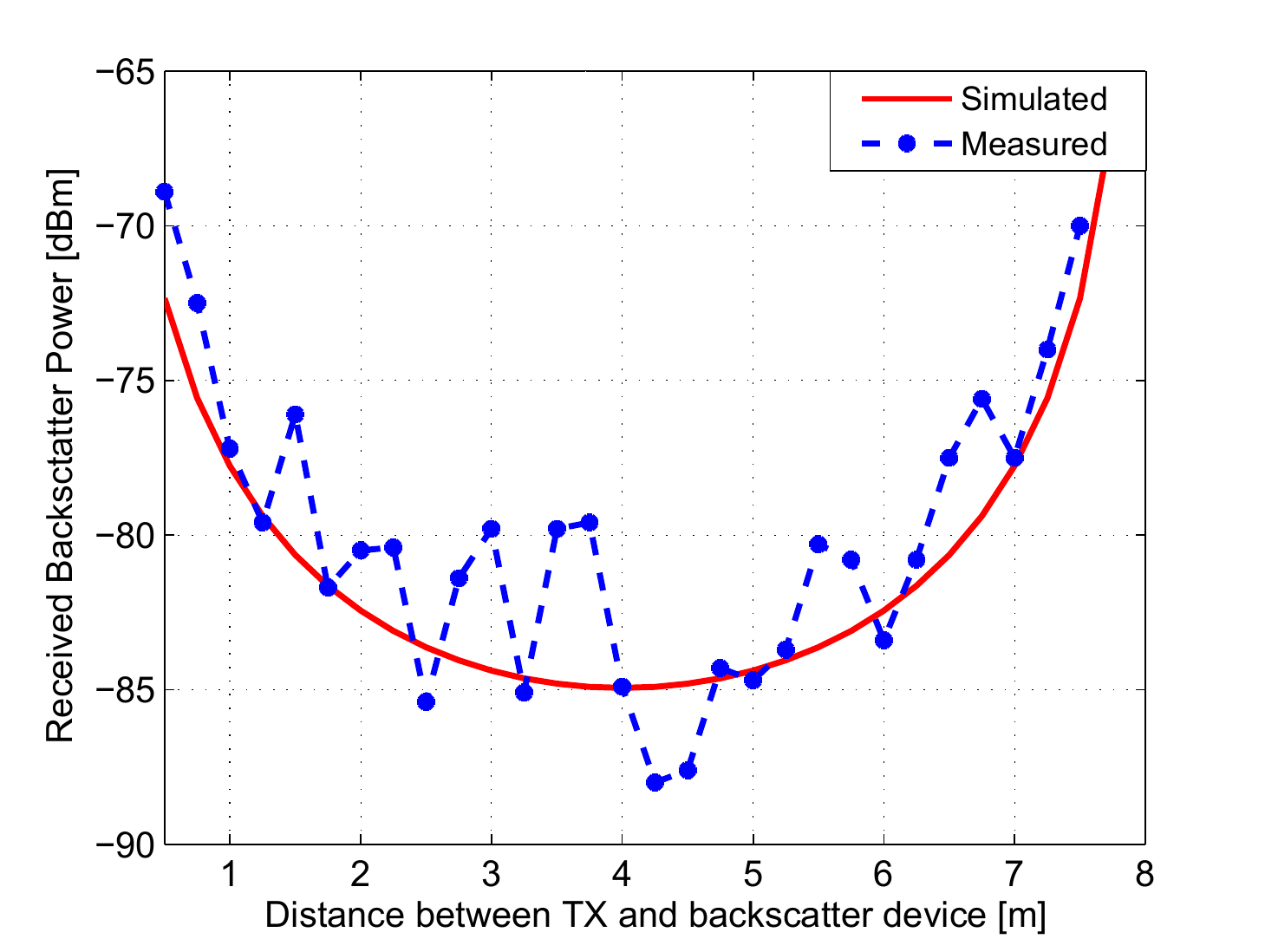} \\
(a) & (b)
\end{tabular}
\caption{Received power against distance between RF transmitter and backscatter device (a) Anechoic chamber (b) Office corridor.}
\label{fig4}
\end{figure*}

\subsection{Results and Discussion}

Fig. \ref{fig4} shows the results for the anechoic chamber and office corridor. The figure shows the received signal strength of the backscatter device against different values of the distance between the backscatter device and RF transmitter. We performed static measurements at an interval of 0.25 distance. {It can be seen that the received backscattered power at the receiver first decreases as the distance between the RF source and backscatter device increases. The received power then increases because the backscatter device now moves closer to the receiver which results in reduced path loss.} Similar trends can be observed for office corridor, however, in this case, the overall received power is lower as compared to the anechoic chamber. This reduction in the received power is due to the corridor environment which consists of glass doors, reflecting walls, smooth ground surface, and wooden doors. The richness of the multipath environment intensifies the impact of fading which then results in loss of power.   

\section{Future research opportunities}

Our in-house designed backscatter device is one of many examples that can be applied in healthcare to reap long-term benefits. However, there are still several research opportunities in the domain of backscatter communications. With this intent, we highlight future research opportunities for applying backscatter communications in the healthcare domain.

\subsection{Improved Testing}

Although a few studies have been conducted where backscatter communication is considered for the in-body implant, no study has been found which considers its impact on living things. {On the contrary, the phantom and dead animal tissues are used for performance evaluation of the network.} While experiments and analysis have been found correct and are able to emulate living tissues, it is more practical and worthwhile to experiment with living animals. In this way, the realistic impact of mobility and temperature on the performance of backscatter devices can be identified, in a more precise manner.
 
\subsection{Multi-hop communication}

{Direct communication between the backscatter device and the receiver can be a limiting factor from the coverage point of view of healthcare services. However, by including multiple intermediate backscatter devices, one can form a chain of reflectors that can act as passive intermediate relays. This can have more longstanding implications for on-body sensors and other wearables. Moreover, to achieve higher levels of reliability in end-to-end healthcare systems, the finite blocklength communication technique can be combined with this mechanism. It would also help in communicating over long ranges without consuming a significant amount of energy of a particular sensor.}

\subsection{Dynamic Frequency shifting}

Although our in-house designed backscatter device works effectively over several meters, the frequency shifting implementation is static. This hinders the performance of the backscatter device when communicating in varying channel conditions. It is, therefore, a desirable characteristic of backscatter systems in healthcare to adapt according to unreliable and changing channel states. For starters, one can focus on dynamic frequency shifting as it would require considerable hardware modifications like oscillators, circuits with multiple clocks, and channel quality detectors.

\subsection{Delay tolerance}

Latency is one of the key performance metrics for wireless devices. Beyond conventional wireless communications, latency is also an important determinant in the provisioning of healthcare services. Most of the time, the delay is not an option in the healthcare domain. {Especially, for critical applications like on-body sensors and miniature implants, latency minimization becomes a critical quality of service requirement.} In this regard, there is a large gap with respect to delay sensitive applications of backscatter communications in healthcare. {Thus, more research focus must be directed towards evaluating and then minimizing the end-to-end communication delays in backscatter communications for healthcare.} 

\subsection{Integration with side channel information}

Typically, medical applications work cooperatively by integrating information from side channels. Existing studies on backscatter communication do not take into account this aspect. For instance, results of MRI scans can be used along with backscatter sensors to identify the cause of the problem in the human body. This integration of side channel can, thus, help in building multi-modal input data which may benefit specific healthcare applications.

\subsection{Dynamic routing mechanism}

Fixed routing is best suitable for static conditions. However, since the backscatter devices in healthcare are expected to work as on-body sensors or in-body implants, they are likely to be mobile for most of the time. A possible solution may be to use control signals to update the number of hops during an iteration. Localization techniques for backscatter communications can also be helpful in designing these dynamic routing messages.

\subsection{Security}

Security is a key aspect of any wireless service. However, no recent work on backscatter communications for healthcare takes into account the inherent vulnerabilities and the broadcast nature of backscatter devices. {Thus, the cost restrictions coupled with the modest computational and storage capabilities of the backscatter sensors motivate the introduction of physical layer security approaches \cite{van2018ambient}. Despite their promising potential to protect from eavesdropping, little has been done to study the prospective benefits of the physical layer techniques in backscatter systems.} Physical layer security can be helpful in securing the devices against unwanted eavesdropping and can be considerably helpful against active jamming attacks. Healthcare service can benefit by jointly combining the lightweight physical layer security techniques for wireless backscatter communications.  

\section{Conclusion}
%%%%%%%%%%%%%%%%%%%%%%%%%%%%
The backscatter communication is emerging as a promising solution for low-powered reliable communications. However, backscatter communication is significantly different from the solutions investigated for conventional healthcare applications. Through an extensive review of the latest studies, this work has identified some key aspects of backscatter communication and highlighted the interplay of different network parameters and their impact in the healthcare domain. A comprehensive study was conducted for indoor propagation conditions using the in-house designed backscatter device. Link budget analysis shows the promise of backscatter devices for indoor healthcare applications.
%%%%%%%%%%%%%%%%%%%%%%%%%%%%
\section*{Acknowledgment}

%%%%%%%%%%%%%%%%%%%%%%%%%%%%
\ifCLASSOPTIONcaptionsoff
  \newpage
\fi
%%%%%%%%%%%%%%%%%%%%%%%%
\bibliographystyle{IEEEtran}
\bibliography{References}
%%%%%%%%%%%%%%%%%%%%%%%%
\begin{IEEEbiographynophoto}{Furqan Jameel}
received his BS in Electrical Engineering (under ICT R\&D funded Program) in 2013 from the Lahore Campus of COMSATS Institute of Information Technology (CIIT), Pakistan. In 2017, he received his Master's degree in Electrical Engineering (funded by prestigious Higher Education Commission Scholarship) at the Islamabad Campus of CIIT. In 2018, he visited Simula Research Laboratory, Oslo, Norway. Currently, he is a researcher at the University of Jyv\"askyl\"a, Finland. His research interests include modeling and performance enhancement of vehicular networks, physical layer security, ambient backscatter communications, and wireless power transfer. He was a recipient of the Outstanding Reviewer Award in 2017 from Elsevier.
\end{IEEEbiographynophoto}
%%%%%%%%%%%%%%%%%%%%%%%%

%%%%%%%%%%%%%%%%%%%%%%%%
\begin{IEEEbiographynophoto}{Ruifeng Duan}
received the B.Eng. in electrical engineering from Hefei University of Technology, China, the M.Sc.(tech.) and D.Sc.(tech.) degrees both in
telecommunications engineering from University of Vaasa, Finland, in 2008 and 2014, respectively. Since August 2014, he has been with the
Department of Communications and Networking, Aalto University, Finland. His current research interests include random matrix theory, free probability theory, extreme value theory, backscatter communications, cognitive radio, ultra-reliable communications, and radio resource
management in wireless systems.
\end{IEEEbiographynophoto}
%%%%%%%%%%%%%%%%%%%%%%%%
\begin{IEEEbiographynophoto}{Zheng Chang} 
(S'10 - M'13 - SM'17) received  the Ph.D. degree from the University of Jyv\"askyl\"a, Finland, in 2013.  He was a Visiting Researcher with Tsinghua University, China, in 2013, and the University of Houston, TX, USA, in 2015. He has been honored by the Ulla Tuominen Foundation, the Nokia Foundation, and the Riitta and Jorma J. Takanen Foundation, the Jorma Ollila Grant for his research excellence. He has served as an editor or guest editor for some IEEE journals and magazines, and has received  best paper awards from IEEE TCGCC and APCC in 2017. He is currently an Assistant Professor with University of Jyv\"askyl\"a. His research interests include vehicular networks, green communications, IoT, cloud/edge computing, and security and privacy.
\end{IEEEbiographynophoto}
%%%%%%%%%%%%%%%%%%%%%%%%%
%%%%%%%%%%%%%%%%%%%%%%%%
\begin{IEEEbiographynophoto}{Aleksi Liljemark}
is with the Department of Communications and Networking, Aalto University, Finland. His current research interests include backscatter communications, and radio resource management in wireless systems.
\end{IEEEbiographynophoto}
%%%%%%%%%%%%%%%%%%%%%%%%%
\begin{IEEEbiographynophoto}{Tapani Ristaniemi} received the M.Sc. degree in mathematics in 1995, the Ph.Lic. degree in applied mathematics in 1997, and the Ph.D. in wireless communications in 2000 from the University of Jyväskylä, Jyvaskyla, Finland. In 2001, he was appointed as a Professor with the Department of Mathematical Information Technology, University of Jyvaskyla. In 2004, he moved to
the Department of Communications Engineering, Tampere University of Technology, Tampere, Finland,
where he was appointed as a Professor in wireless communications. Prof. Ristaniemi is currently a Consultant and a member of the Board of Directors of Magister Solutions Ltd. He is currently an Editorial Board Member of Wireless Networks and International Journal of Communication Systems.
\end{IEEEbiographynophoto}
%%%%%%%%%%%%%%%%%%%%%%%%

%%%%%%%%%%%%%%%%%%%%%%%%%
\begin{IEEEbiographynophoto}{Riku Jantti} is a professor (tenured) in Communications Engineering and the head of the Department of Communications and Networking at Aalto University's School of Electrical Engineering, Finland. He received his M.Sc
(with distinction) in electrical engineering in 1997, and the D.Sc. (with distinction) in automation and systems technology in 2001, both from Helsinki University of Technology (TKK). The research interests of Prof. Jantti include radio resource control and optimization for machine type communications, cloud based radio access networks, spectrum and co-existence management, and RF inference.
\end{IEEEbiographynophoto}
%%%%%%%%%%%%%%%%%%%%%%%%

\end{document}